# THE IMPORTANCE OF EMOTIONAL INTELLIGENCE IN LEADERSHIP FOR BUILDING AN EFFECTIVE TEAM

Joanna ĆWIĄKAŁA[1], Waldemar GAJDA[2], Michał ĆWIĄKAŁA[3], Ernest GÓRKA[4],
Dariusz BARAN[5], Gabriela WOJAK[6], Piotr MRZYGŁÓD[7], Maciej FRASUNKIEWICZ[8],
Piotr RĘCZAJSKI[9], Jan PIWNIK[10]

[1] I'M Brand Institute sp. z o.o.; j.cwiakala@imbrandinstitute.pl, ORCID: 0000-0002-4269-9492
[2] Warszawska Szkoła Zarządzania-Szkoła Wyższa; waldemar.gajda@wsz-sw.edu.pl,
ORCID: 0000-0003-0739-4340
[3] Wyższa Szkoła Kształcenia Zawodowego; michal.cwiakala@wskz.pl, ORCID: 0000-0001-9706-864X
[4] WSB - National-Louis University, College of Social and Computer Sciences; ewgorka@wsb-nlu.edu.pl,
ORCID: 0009-0006-3293-5670
[5] WSB - National-Louis University, College of Social and Computer Sciences; dkbaran@wsb-nlu.edu.pl,
ORCID: 0009-0006-8697-5459
[6] Jagiellonian University, Faculty of Management and Social Communication; gabriela.wojak@student.uj.edu.pl,
ORCID: 0009-0003-2958-365X
[7] Piotr Mrzygłód Sprzedaż-Marketing-Consulting; piotr@marketing-sprzedaz.pl, ORCID: 0009-0006-5269-0359
[8] F3-TFS sp. z o.o.; m.frasunkiewicz@imbrandinstitute.pl, ORCID: 0009-0006-6079-4924
[9] MAMASTUDIO Pawlik, Ręczajski, spółka jawna; piotr@mamastudio.pl, ORCID: 0009-0000-4745-5940
[10] WSB Merito University in Gdańsk, Faculty of Computer Science and New Technologies;
jpiwnik@wsb.gda.pl, ORCID: 0000-0001-9436-7142
* Correspondence author

**Purpose:** This study investigates the role of emotional intelligence (EI) in leadership and its impact on building effective teams. It explores how EI competencies influence team collaboration, motivation, and conflict resolution within organizations.

**Design/methodology/approach:** A survey of 100 professionals assessed leadership traits and EI impact on teams. Statistical analysis identified key correlations.

**Findings:** Leaders with high EI are more empathetic, ethical, and effective in motivating teams. Strong correlations were found between EI, social competence, and leadership success.

**Research limitations/implications:** The study is limited to a specific demographic. Future research should explore diverse industries and cultural contexts.

**Practical implications:** Developing EI in leadership training enhances team cohesion, reduces conflicts, and improves motivation.

**Social implications:** EI-driven leadership fosters healthier workplace cultures, reducing stress and enhancing job satisfaction.

**Originality/value:** This study provides empirical insights into EI's role in leadership, filling gaps in research by linking EI with leadership effectiveness.

**Category of the paper:** research paper.

# 1. Introduction

In today's fast-evolving business landscape, leadership is no longer defined solely by technical expertise or strategic decision-making. The ability to manage emotions, build strong interpersonal relationships, and foster a positive team environment has become essential for effective leadership. Emotional intelligence (EI), which encompasses self-awareness, self-regulation, motivation, empathy, and social skills, has gained significant recognition as a key determinant of leadership success.

The modern workplace presents numerous challenges, including managing diverse teams, navigating organizational change, and resolving conflicts efficiently. Leaders who exhibit high EI are better equipped to handle these complexities by understanding their own emotions and those of their colleagues. This capacity allows them to build trust, communicate effectively, and inspire team members to achieve common goals. Research has demonstrated that leaders with strong EI create environments where employees feel valued, motivated, and engaged, ultimately leading to higher productivity and organizational success.

Traditional leadership models often prioritized authority, control, and decision-making prowess. However, contemporary perspectives emphasize the importance of interpersonal effectiveness and emotional intelligence in fostering a resilient and adaptive workforce. Leaders with high EI can navigate crises with composure, encourage innovation by fostering psychological safety, and strengthen workplace relationships through empathy and emotional awareness. These factors contribute to a more cohesive, motivated, and high-performing team.

Additionally, emotional intelligence plays a crucial role in conflict resolution. Workplaces are often prone to misunderstandings, differing perspectives, and interpersonal tensions. Leaders with strong EI can de-escalate conflicts, mediate disputes, and create solutions that promote harmony and cooperation. Their ability to regulate their own emotions and respond thoughtfully rather than react impulsively ensures a constructive and balanced approach to leadership.

This study aims to explore the impact of EI on leadership effectiveness and its role in enhancing team dynamics, motivation, and organizational well-being. By analysing survey data from professionals across various industries, this research provides empirical insights into how EI influences leadership outcomes. The findings will inform strategies for integrating EI development into leadership training programs, ensuring that organizations cultivate leaders who can drive success through emotional and social competence. As businesses continue to evolve, understanding and applying emotional intelligence in leadership will remain crucial for fostering inclusive, high-performing workplaces. Organizations that fail to recognize the importance of EI risk falling behind in the ever-changing corporate environment. Thus, embracing EI-driven leadership strategies is not just beneficial but essential for sustained success and innovation.

## 2. The evolution and models of emotional intelligence

Over the past three decades, emotional intelligence (EI) has evolved from a relatively obscure scientific concept to a widely recognized and extensively discussed construct in both academic and practical contexts. Initially facing challenges in conceptualization and measurement, EI has endured skepticism and criticism to become a flourishing field of research. Despite early concerns regarding its definition and assessment, the domain of emotional intelligence continues to advance. Ongoing efforts aim to refine its components and measurement tools, incorporating insights from emotion theory and related disciplines to develop innovative approaches for assessing emotional intelligence (Mortillaro, Schlegel, 2023). These approaches are grounded in foundational scientific literature on emotions and appraisal theories, ensuring a more comprehensive understanding of EI within contemporary research frameworks.

Emotional intelligence is a multifaceted concept that has gained increasing significance across various disciplines. The core components of EI include self-awareness, self-regulation, empathy, and social skills (Miao, Humphrey, Qian, 2017). Self-awareness entails recognizing and understanding one's own emotions and their influence on thoughts and behaviors. Self-regulation refers to effectively managing emotions and adapting to different situations without becoming overwhelmed. Empathy is the ability to understand and share the feelings of others, facilitating improved interpersonal relationships. Social skills encompass effective communication, conflict resolution, and teamwork abilities, all of which are crucial for successful interactions in both personal and professional contexts (Walter, Humphrey, Cole, 2012). Collectively, these components contribute to an individual's overall emotional intelligence and play a vital role in personal and professional success.

Different researchers have provided various definitions of emotional intelligence, each emphasizing different aspects of its application. Salovey and Mayer (1990) described EI as the ability to monitor and differentiate emotions in oneself and others, using this information to guide thoughts and actions. Goleman (1995) expanded on this by incorporating interpersonal skills such as self-motivation, relationship management, and empathy, which are crucial for success in professional and personal settings. Bar-On (1997) focused on EI as a set of competencies and skills necessary for coping with environmental demands, highlighting adaptability as a key feature. Meanwhile, Caruso and Salovey (2004) stressed emotional regulation and awareness as essential for intelligent and adaptive behaviour. Barsade and Gibson (2007) underscored the importance of EI in workplace performance, emphasizing the ability to manage emotions effectively to foster collaboration and efficiency. These varying perspectives reflect the broad scope of EI, demonstrating its significance in both personal development and organizational success.

In the academic literature, two main models of emotional intelligence have been distinguished: the mixed model (popular approach) and the ability model (scientific approach).

The mixed model, also referred to as the popular model of emotional intelligence, was popularized by Daniel Goleman in his 1995 book Emotional Intelligence (Goleman, 1995). Goleman defines EI as a combination of traits and skills encompassing both emotional and social aspects. This model integrates elements such as self-awareness, emotional regulation, self-motivation, empathy, and social skills. Reuven Bar-On further refined this perspective by distinguishing five key functional areas: intrapersonal skills (e.g., assertiveness, self-awareness, independence), interpersonal skills (e.g., empathy, relationship management), adaptability (e.g., flexibility, problem-solving), stress management (e.g., emotional control, stress resilience), and general mood (e.g., optimism, happiness) (Sadowska, Brachowicz, 2008). While widely used in management and personal development contexts, the mixed model has been criticized for its broad and imprecise definition, as it combines various traits rather than focusing solely on cognitive abilities.

The ability model, often regarded as the scientific model of emotional intelligence, was introduced by John D. Mayer and Peter Salovey in 1990. This model focuses exclusively on cognitive abilities related to emotions, viewing emotions as structured responses that operate through multiple psychological subsystems, including physiological, cognitive, and motivational components (Mayer, Salovey, 1990).

Mayer and Salovey (1990) define EI as the ability to perceive emotions in oneself and others, use emotions to support cognition and problem-solving, understand their complexities and evolution, and regulate emotions to foster intellectual and emotional growth.

These abilities are key determinants of emotional intelligence. The ability model is regarded as more scientific and precise since it is based on measurable cognitive functions and has been extensively validated through empirical research. It has been applied in various contexts, including education, mental health, and workplace performance (Sadowska, Brachowicz, 2008). Additionally, Czesław S. Nosal (1998) proposed that EI consists of two key psychological domains: intellectual and emotional. He argued that emotional intelligence results from the interaction of mechanisms responsible for the generation, categorization, and interpretation of emotions, which influence emotional states and cognitive processing.

Unlike Mayer and Salovey's ability model, Goleman's mixed model expands on the fundamental abilities of perceiving, using, understanding, and managing emotions by incorporating additional elements such as social skills, motivation, and self-regulation (Salovey, Grewal, 2005). This broader perspective highlights the importance of interpersonal competencies in emotional intelligence. Bar-On's framework further emphasizes personal and social competencies over specific cognitive functions related to emotion processing. The mixed model thus underlines the integration of emotional awareness with social proficiency and self-regulation, positioning it as a practical approach to emotional intelligence in professional and personal development (Salovey, Grewal, 2005).

## 3. Emotional intelligence in management

Emotional intelligence plays a crucial role in effective workplace performance and organizational success. Goleman (1999) emphasized that EI is fundamental to professional achievement, arguing that managing one's emotions and recognizing the emotions of others contribute significantly to leadership and teamwork. While general intelligence remains important, EI has been identified as a key factor in distinguishing high-performing employees and leaders. Bradberry and Greaves (2006) noted that although some individuals achieve success with lower EI levels, their accomplishments often stem from high general intelligence and strong self-discipline.

Recent studies highlight the growing relevance of EI in professional settings. Research by Matczak and Knopp (2013) suggests that individuals with higher EI levels perform better in job interviews, exhibit greater leadership effectiveness, and achieve higher career success. Further empirical studies reinforce these findings. Skwarek's research on professional athletes demonstrated that those with higher EI scores excel in competitive environments due to their superior emotional regulation and resilience (Matczak, Knopp, 2013). Additionally, studies by Basińska, Jaskólska, and Piórowski (2007) on military personnel revealed that emotionally intelligent soldiers display higher job satisfaction and career ambition, while those with lower EI levels are more susceptible to occupational burnout.

From an organizational perspective, EI is essential for managing team dynamics and improving workplace culture. Goleman (1999) argued that EI-driven leadership enhances team productivity, fosters positive communication, and mitigates workplace conflicts. Modern companies increasingly seek employees with strong interpersonal skills, adaptability under pressure, and self-motivation. The ability to recognize, understand, and manage emotions contributes to building high-performing teams and sustainable business growth. Consequently, organizations that invest in developing EI competencies among employees gain a competitive advantage by enhancing collaboration, employee engagement, and overall job performance.

Recent research further deepens the understanding of how emotional intelligence (EI) competencies contribute to organizational effectiveness. Iyer (2024) conducted a comprehensive study examining the impact of self-awareness, self-management, social awareness, and relationship management on key leadership outcomes such as decision-making, team performance, and employee engagement. The findings revealed that emotionally intelligent leaders are better equipped to navigate complex organizational dynamics, foster trust, and resolve conflicts effectively. Moreover, the study emphasized the mediating role of transformational leadership behaviors and social exchange mechanisms, such as mutual respect and trust-building, in enhancing leadership outcomes. Leaders with high EI were shown to inspire commitment, drive team cohesion, and improve organizational performance through emotionally attuned interactions.

Recent findings by López González et al. (2024) confirm a strong positive correlation between emotional intelligence and leadership competencies among university students, particularly highlighting the predictive role of the "use of emotions" dimension. The study demonstrates that specific components of emotional intelligence can reliably forecast leadership potential, suggesting that developing emotional regulation and expression skills is vital for cultivating future leaders in higher education contexts.

Further supporting the critical role of emotional intelligence in leadership, Nwagwu and Henry (2025) emphasize that emotionally intelligent leaders are more capable of regulating their own emotions and understanding the emotional states of their teams. This emotional atonement enables them to build stronger relationships, communicate effectively, and make more informed decisions in high-pressure environments. The study highlights that high EI levels among leaders contribute directly to improved leadership effectiveness by fostering trust, motivation, and team alignment. As a result, organizations are encouraged to integrate emotional intelligence training into leadership development programs, recognizing its value in enhancing both individual and organizational performance.

## 4. Research material and methodology

The survey questionnaire consisted of 21 questions. The questions were designed to test respondents' knowledge of a leader's emotional intelligence and to assess the impact of emotional intelligence on management.

The research was conducted on a social network via Google Forms, where a digital version of the questionnaire was prepared. The developed survey tool received 100 responses from randomly selected respondents who decided to respond to the request to complete the questionnaire.

The following is a list of questions answered by the respondents:

1. Do you know what emotional intelligence is?
   A. Yes.
   B. No.
2. What do you think emotional intelligence is?
   A. The ability to control and manage your emotions in stressful and conflict situations.
   B. The ability to recognise and understand one's own emotions and the emotions of others in a team.
   C. The ability to make decisions based on data analysis and intuition.
3. Do you think your current leader has a high level of emotional intelligence?
   Scale from 1 to 5
   1 – Definitely no.

    2 – No.

    3 – Neutral.

    4 – Yes.

    5 – Definitely yes.

4. What leadership qualities in your experience are the best examples of high emotional intelligence?

    A. Empathy.

    B. Fairness.

    C. Analytical mind.

5. Have you noticed that your leader is able to manage his/her emotions effectively during conflicts?

    A. Yes.

    B. No.

6. How does your leader manage emotions?

    A. Can adapt them to the current situation.

    B. Reacts nervously.

    C. Is calm and does not show emotion.

7. How does your leader resolve conflicts?

    A. Through dialogue.

    B. Through stressful conversations with employees.

    C. By waiting for the conflict to resolve itself.

8. Does your leader demonstrate a high level of empathy with your team?

    Scale from 1 to 5

    1 – Definitely no.

    2 – No.

    3 – Neutral.

    4 – Yes.

    5 – Definitely yes.

9. How has your leader's empathy influenced team collaboration?

    A. It integrated and motivated the team.

    B. The team works without commitment.

    C. Team members are stressed.

10. Do you think your leader acts in accordance with ethical principles?

    A. Yes.

    B. No.

11. What actions or decisions of your leader show his/her compliance with ethical principles?

    A. Treatment of colleagues.

    B. Resolving conflicts within the team.

    C. Respecting others and being respectful.

12. Can your leader effectively motivate the team to achieve goals?

    Scale from 1 to 5

    1 – Definitely no.

    2 – No.

    3 – Neutral.

    4 – Yes.

    5 – Definitely yes.

13. What motivational methods does your leader use?

    A. Recognition and rewards.

    B. Opportunities for professional development.

    C. Creating a friendly workplace.

14. Does your leader have developed social skills?

    Scale from 1 to 5

    1 – Definitely no.

    2 – No.

    3 – Neutral.

    4 – Yes.

    5 – Definitely yes.

15. What personality traits of your leader help to build positive team relationships?

    A. Empathy.

    B. Communicativeness.

    C. Authenticity.

16. How old are you?

    A. 18-26.

    B. 27-35.

    C. 36-45.

    D. 46-55.

    E. 55-59

17. Where do you live?

    A. Village.

    B. City of up to 50 000 inhabitants.

    C. City of 50,000 to 100,000 inhabitants.

    D. City with more than 100,000 inhabitants.

18. How many years' work experience do you have?

    A. Up to 5 years.

    B. Between 5 and 15 years.

    C. Over 15 years.

19. What education do you have?
    A.  Vocational.
    B.  Secondary.
    C.  Higher.
20. What position in the organisation do you hold?
    A.  Serial employee.
    B.  Management position.
21. How many people does your organisation employ?
    A.  Up to 20.
    B.  Between 21 and 200.
    C.  More than 200.

## 4.1. Findings

The survey data was analysed using statistical methods. Once the data was collected, the results were visualized using bar charts and pie charts to highlight key trends and distributions. Bar charts were used to compare the scale of survey responses, while pie charts illustrated proportional relationships in the data sets. The survey data were organized, and all visualizations and calculations were performed in Microsoft Excel.

The study does not strictly follow either the mixed model of emotional intelligence as proposed by Goleman, or the ability model by Mayer and Salovey. However, it draws on elements from both frameworks to assess emotional intelligence in leadership. Goleman's mixed model is more comprehensive in integrating personality traits and leadership qualities, making it a suitable basis for evaluating how emotional intelligence impacts leadership behaviors and team outcomes. Mayer and Salovey's model, on the other hand, offers a more structured approach to measuring emotional competencies. While the study does not solely rely on one specific model, incorporating elements from both frameworks strengthens the theoretical foundation of the research, providing a broader understanding of emotional intelligence and its effects on leadership.

It is important to treat this study as a pilot project due to the relatively small number of respondents. With only 100 participants completing the survey, the sample size is limited, which may impact the generalizability of the findings. A small sample size can also result in a higher margin of error, potentially affecting the reliability and validity of the results. Additionally, the diversity of the respondents while offering some range of perspectives might not fully represent the broader population of professionals across various industries.

Given these limitations, the results of this study should be seen as exploratory and not definitive. The findings provide valuable preliminary insights into the role of emotional intelligence in leadership, but further research with a larger and more diverse sample is needed to confirm these results and draw more robust conclusions. The study's pilot nature also allows for refinements in the research design, which could improve the accuracy and applicability of future investigations into this area.

A collective summary of the research results for all 100 respondents, in accordance with the 15 questions in the survey, is presented in Figures 1 to 22.

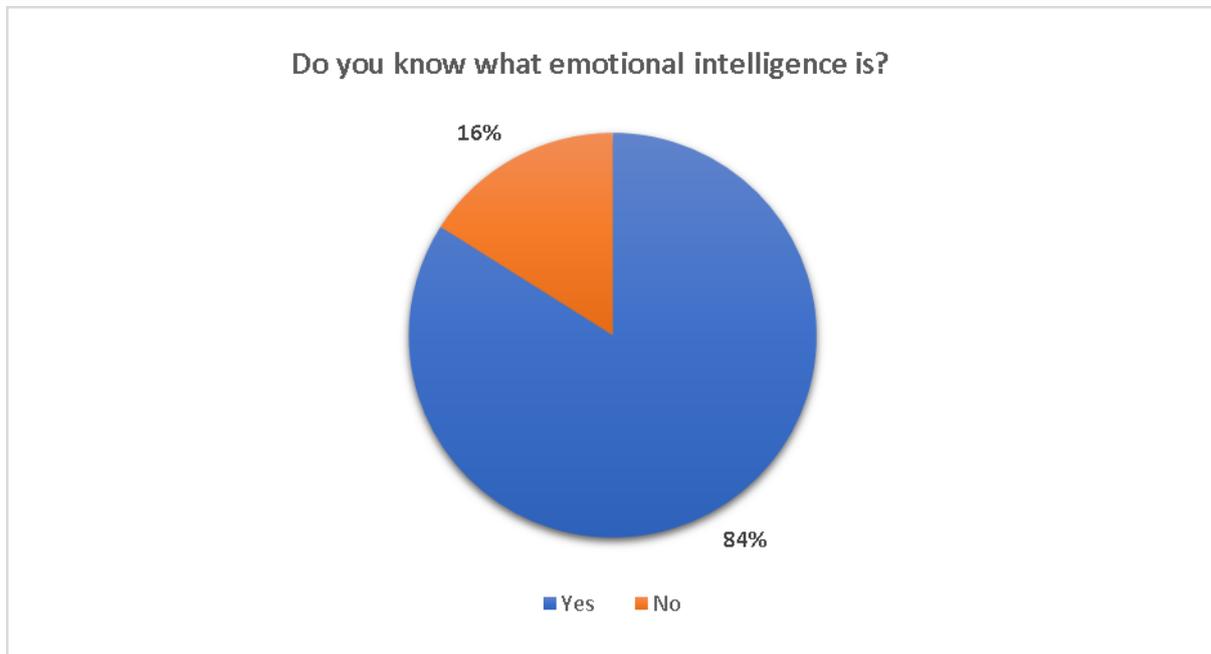

**Figure 1.** Answers from enterprise employees who have experienced a variety of leaders' management styles to question 1 of the research survey.

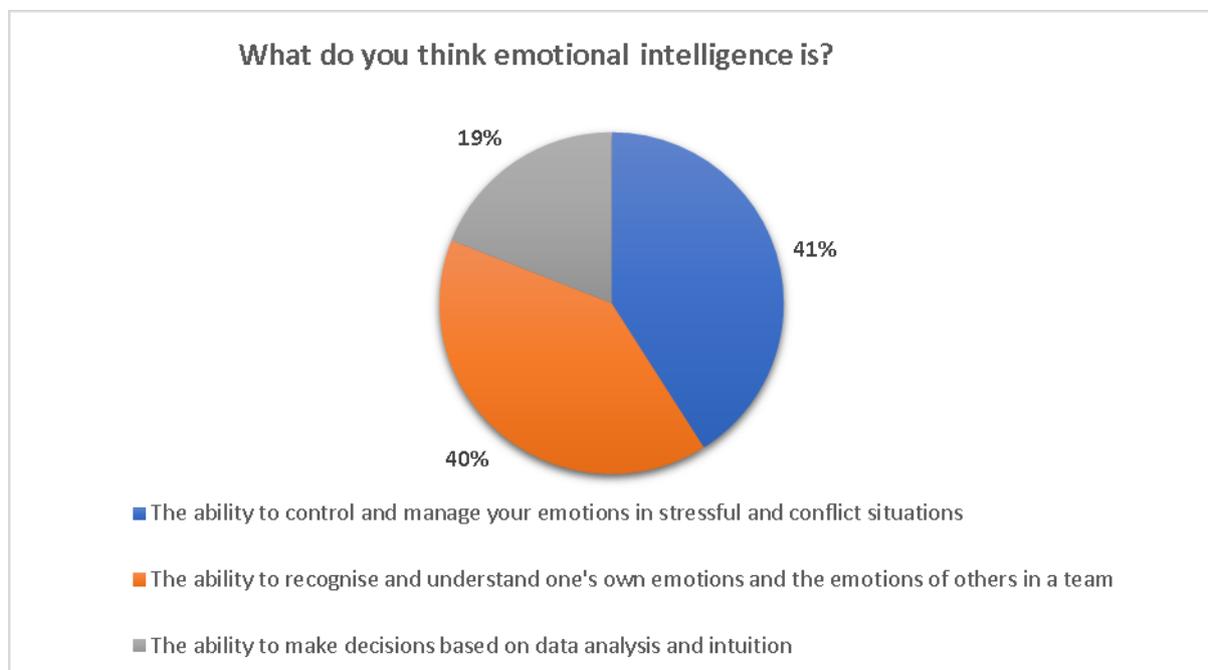

**Figure 2.** Answers from enterprise employees who have experienced a variety of leaders' management styles to question 2 of the research survey.

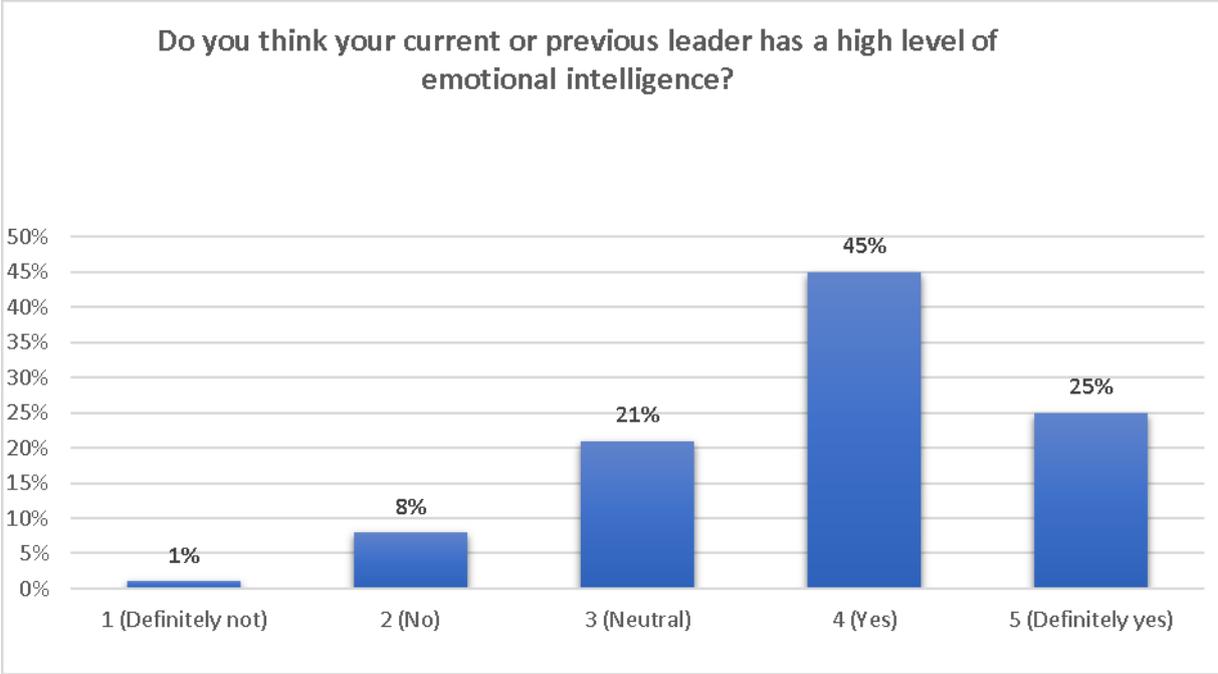

**Figure 3.** Answers from enterprise employees who have experienced a variety of leaders' management styles to question 3 of the research survey.

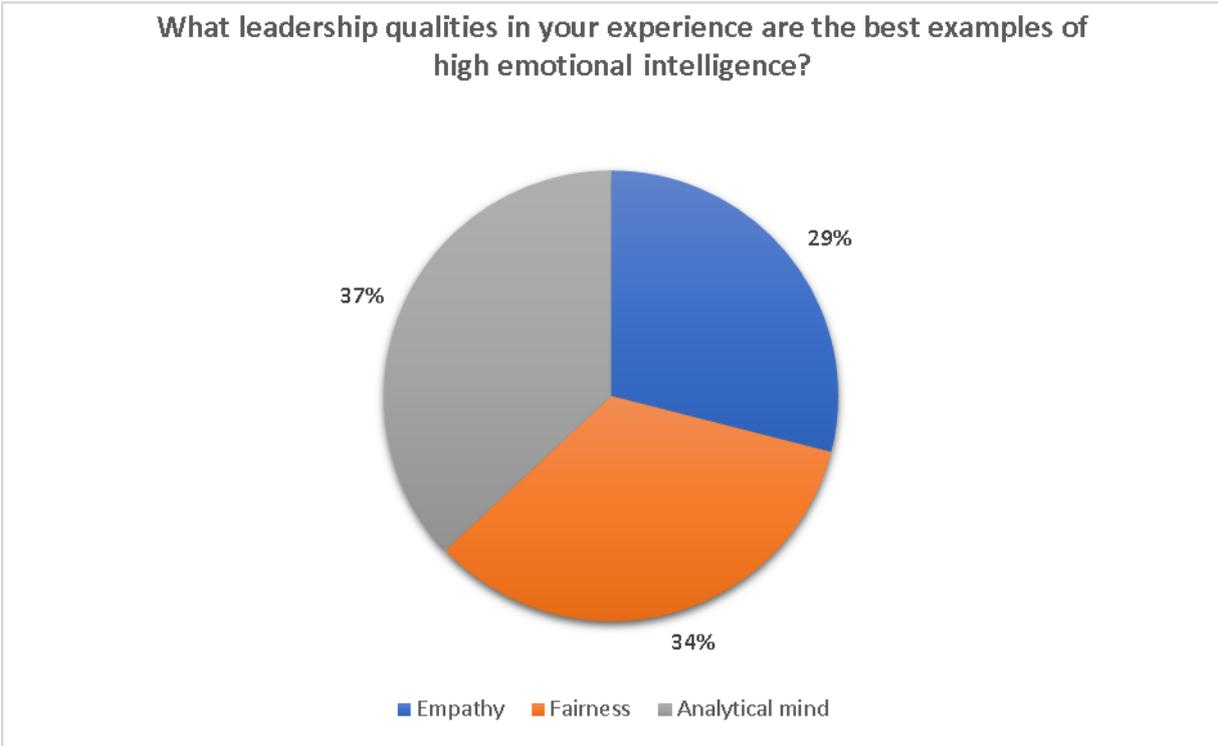

**Figure 4.** Answers from enterprise employees who have experienced a variety of leaders' management styles to question 4 of the research survey.

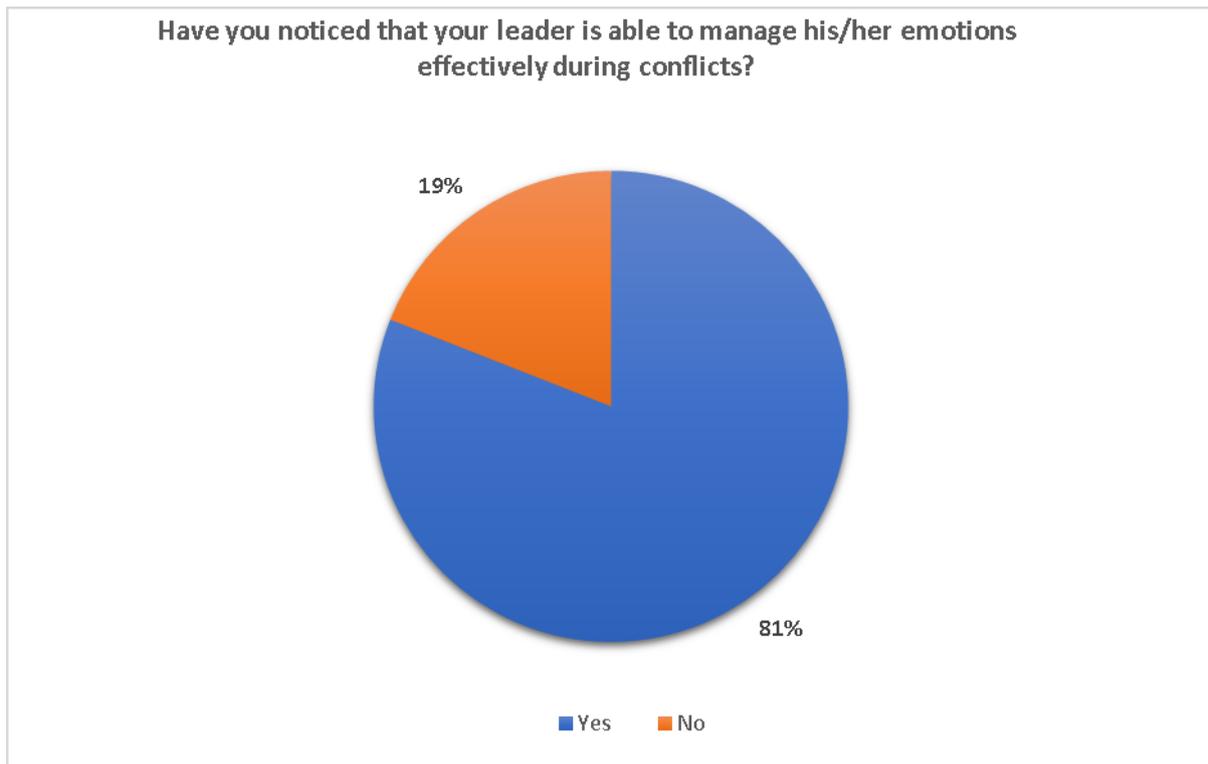

**Figure 5.** Answers from enterprise employees who have experienced a variety of leaders' management styles to question 5 of the research survey.

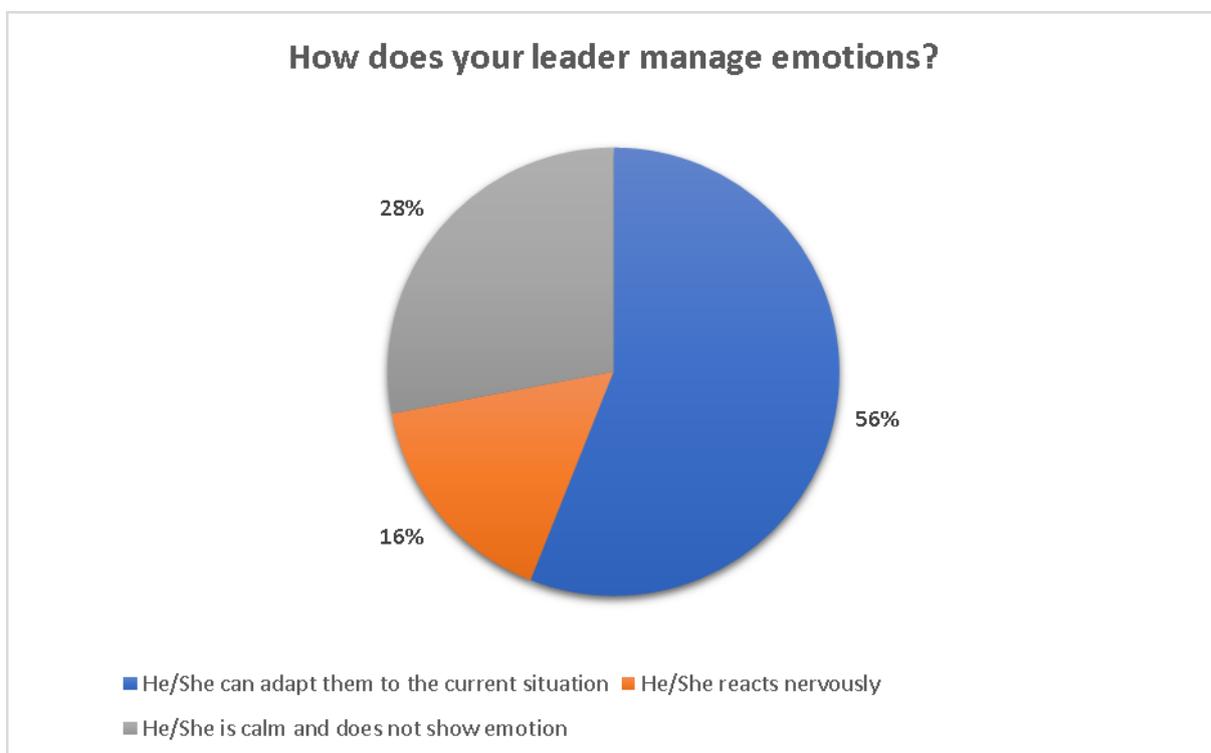

**Figure 6.** Answers from enterprise employees who have experienced a variety of leaders' management styles to question 6 of the research survey.

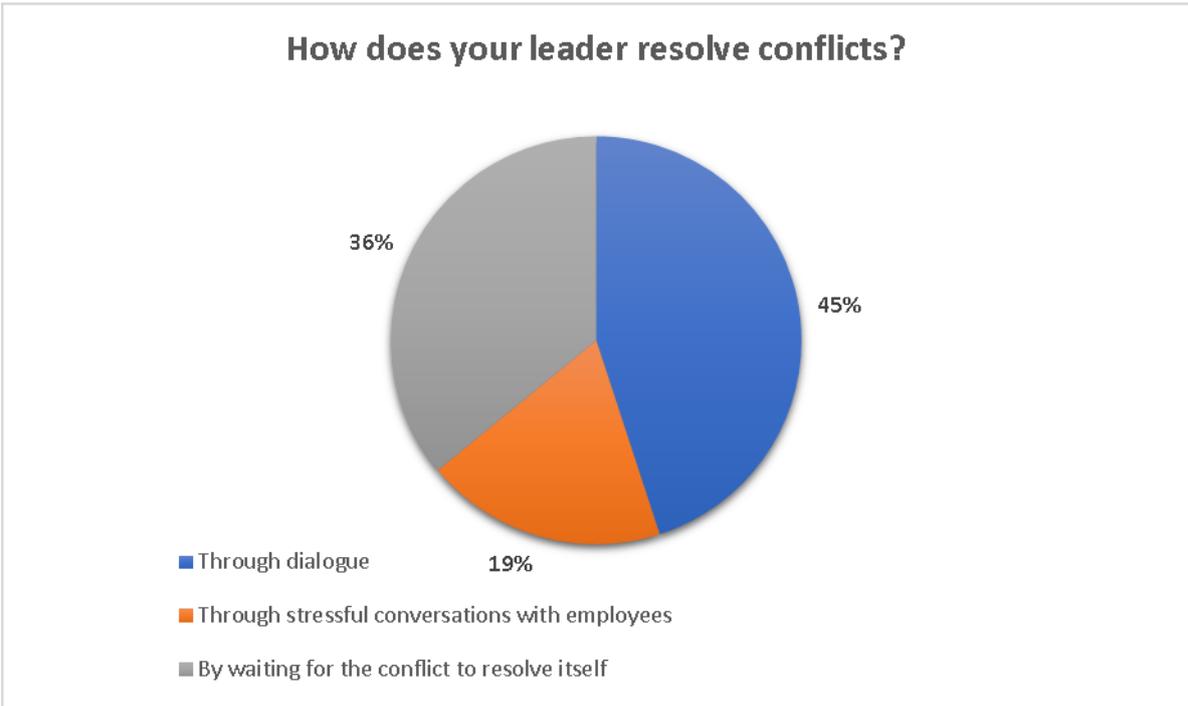

**Figure 7.** Answers from enterprise employees who have experienced a variety of leaders' management styles to question 7 of the research survey.

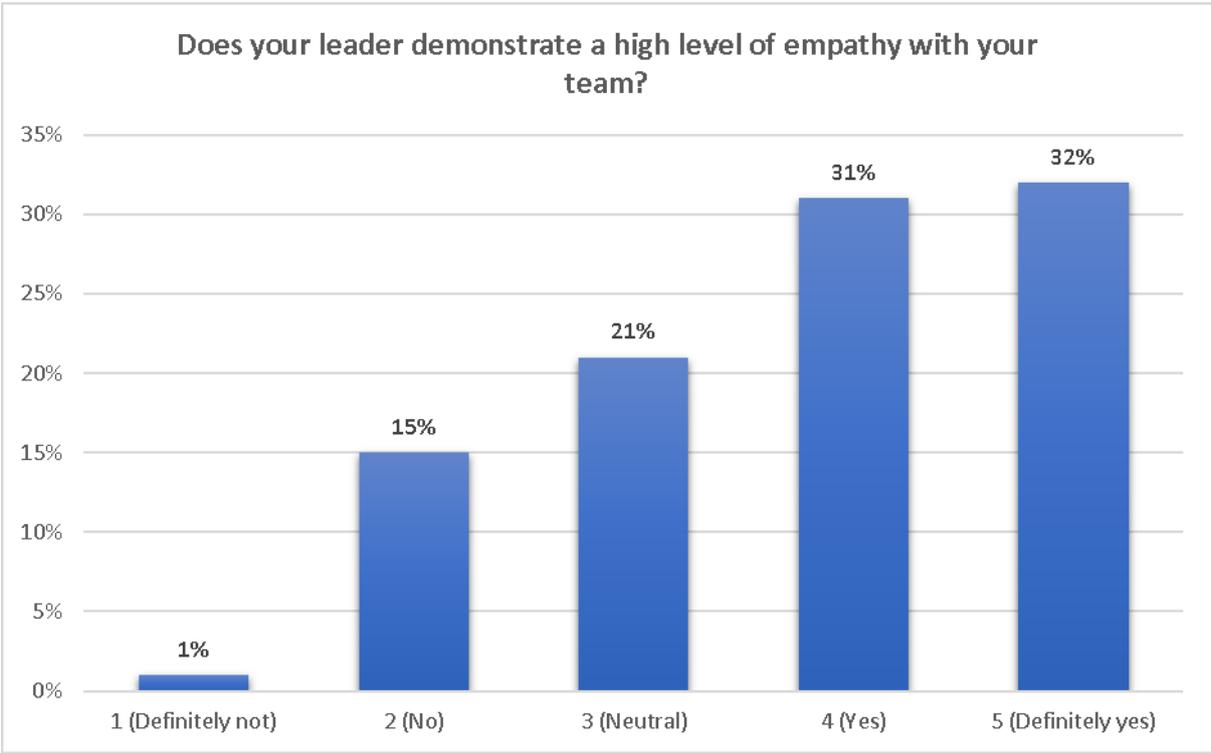

**Figure 8.** Answers from enterprise employees who have experienced a variety of leaders' management styles to question 8 of the research survey.

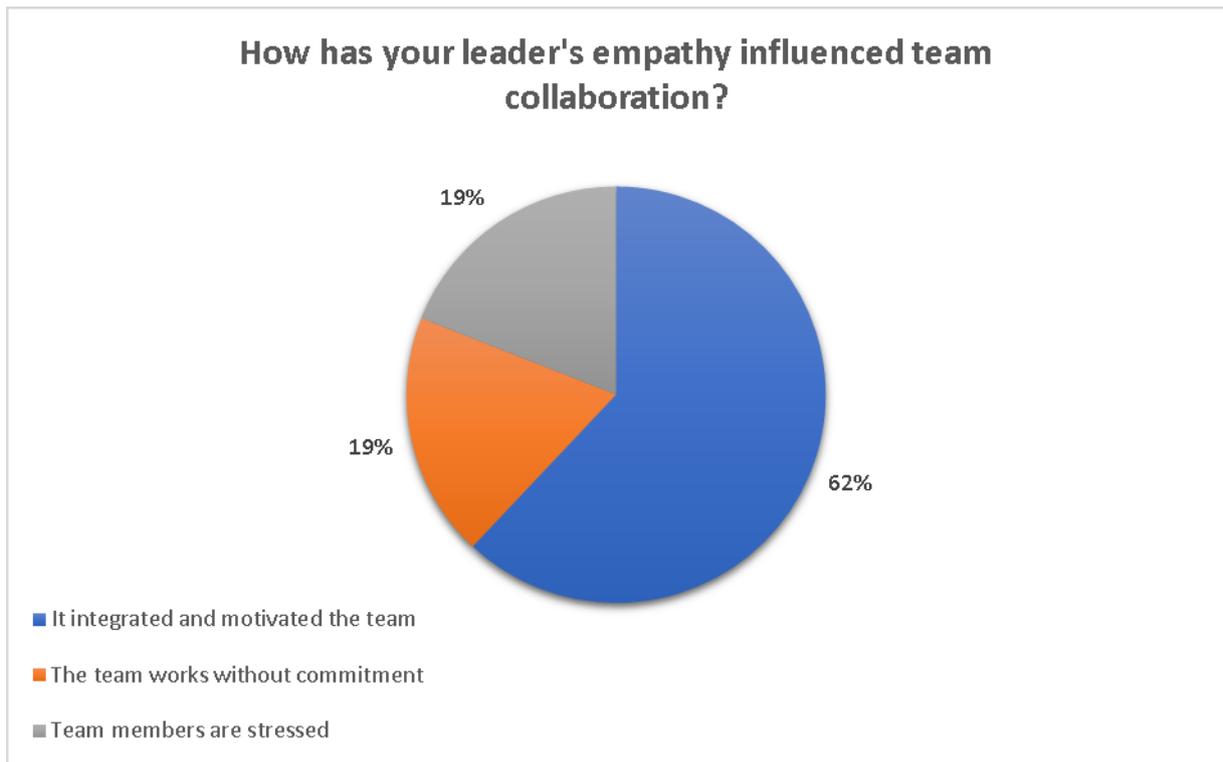

**Figure 9.** Answers from enterprise employees who have experienced a variety of leaders' management styles to question 9 of the research survey.

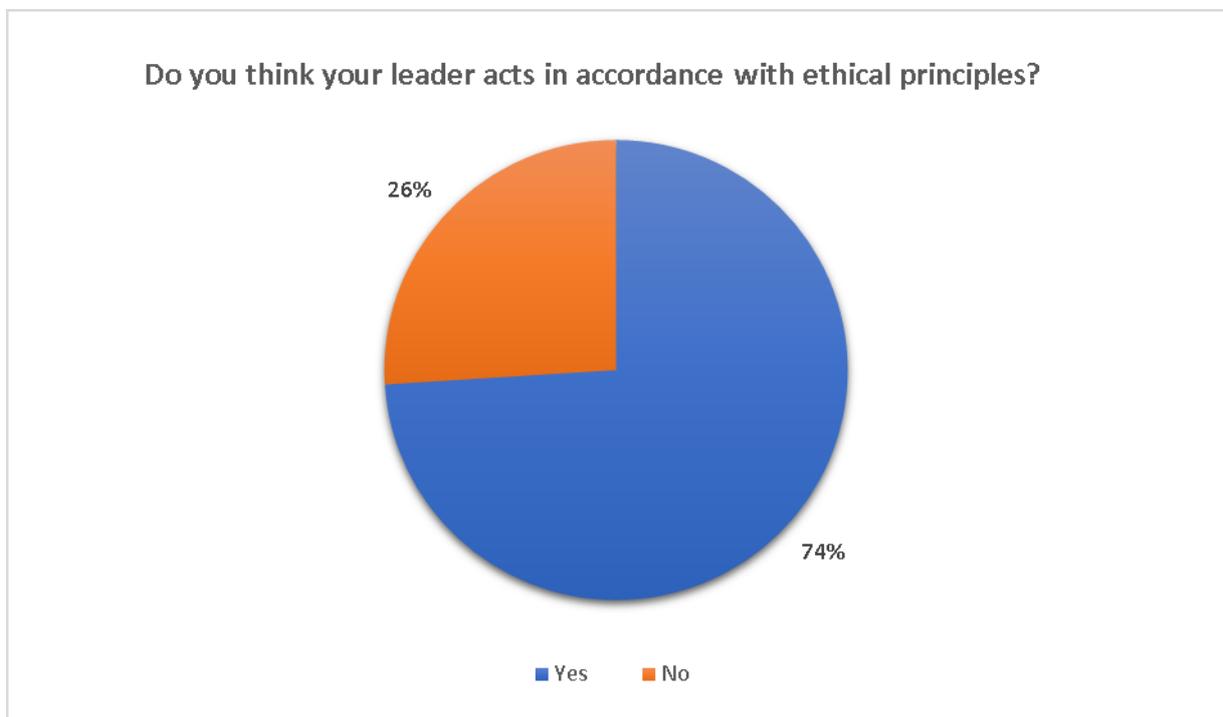

**Figure 10.** Answers from enterprise employees who have experienced a variety of leaders' management styles to question 10 of the research survey.

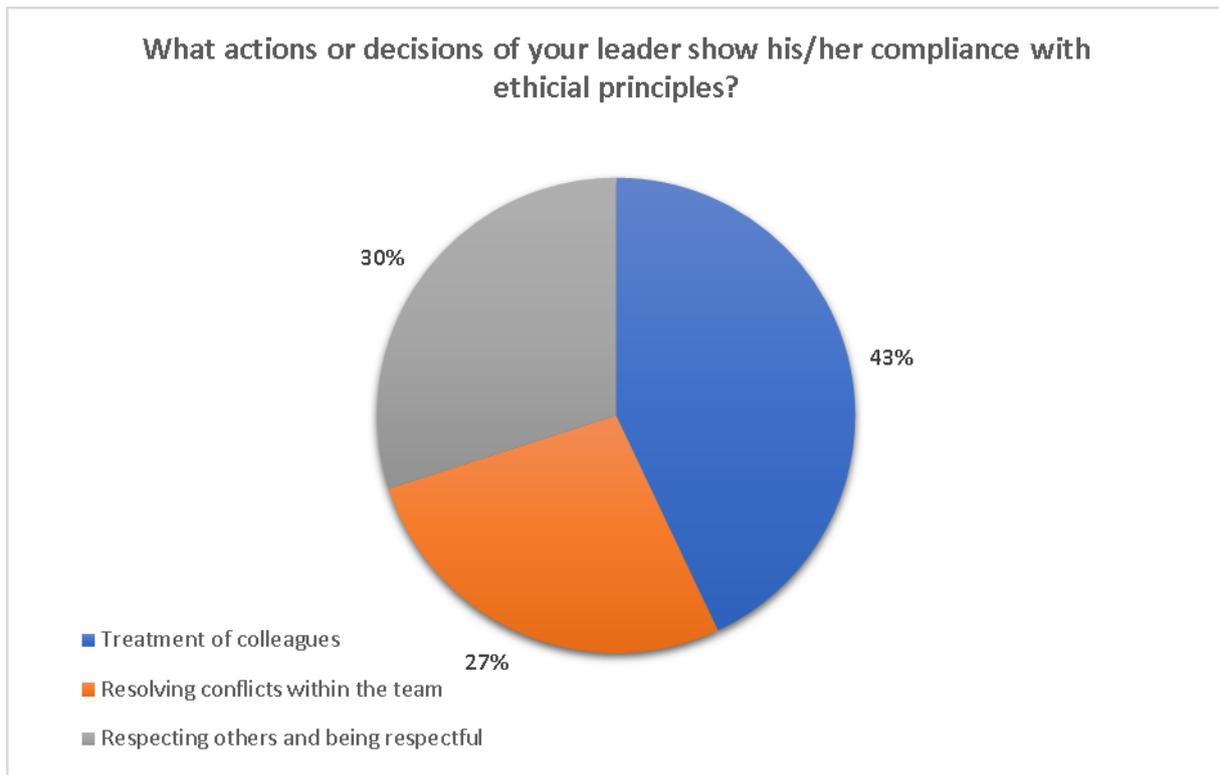

**Figure 11.** Answers from enterprise employees who have experienced a variety of leaders' management styles to question 11 of the research survey.

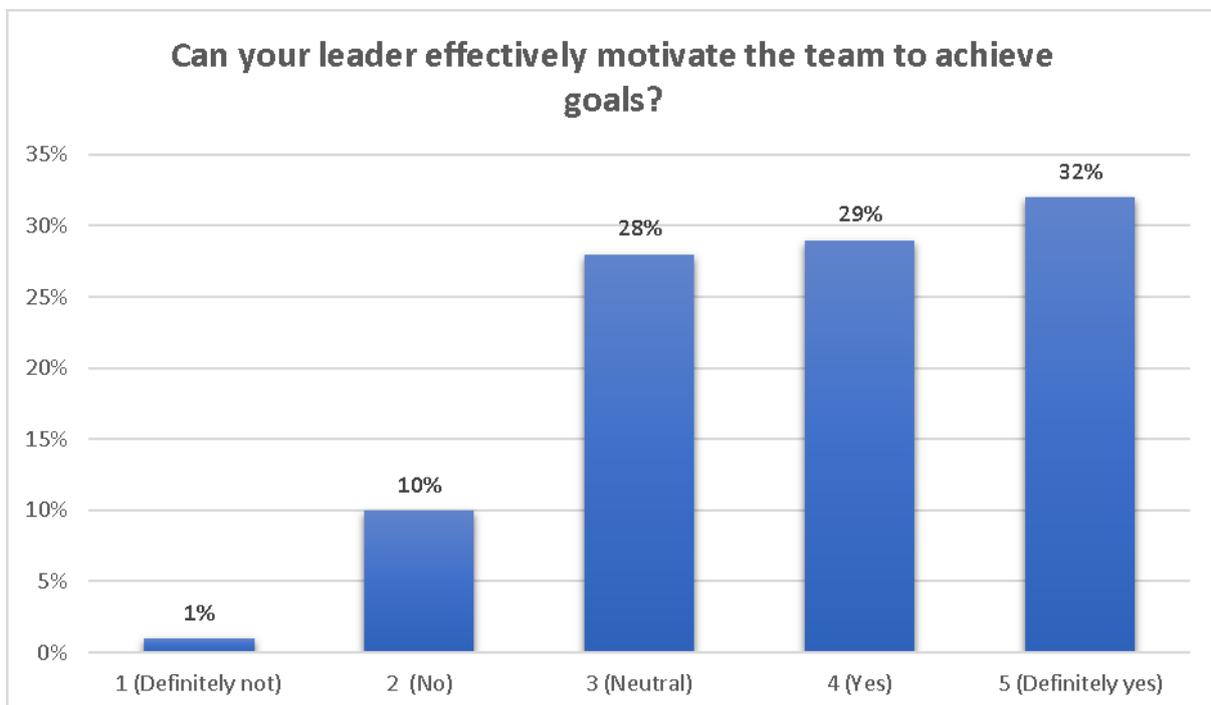

**Figure 12.** Answers from enterprise employees who have experienced a variety of leaders' management styles to question 12 of the research survey.

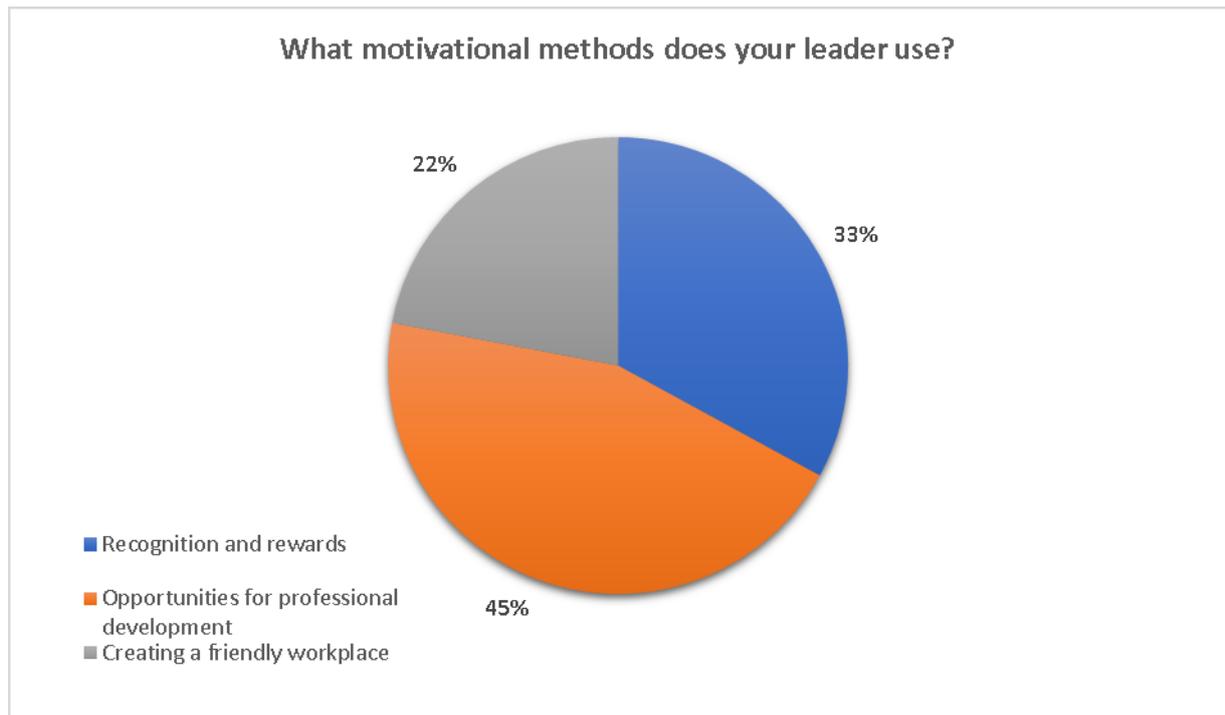

**Figure 13.** Answers from enterprise employees who have experienced a variety of leaders' management styles to question 13 of the research survey.

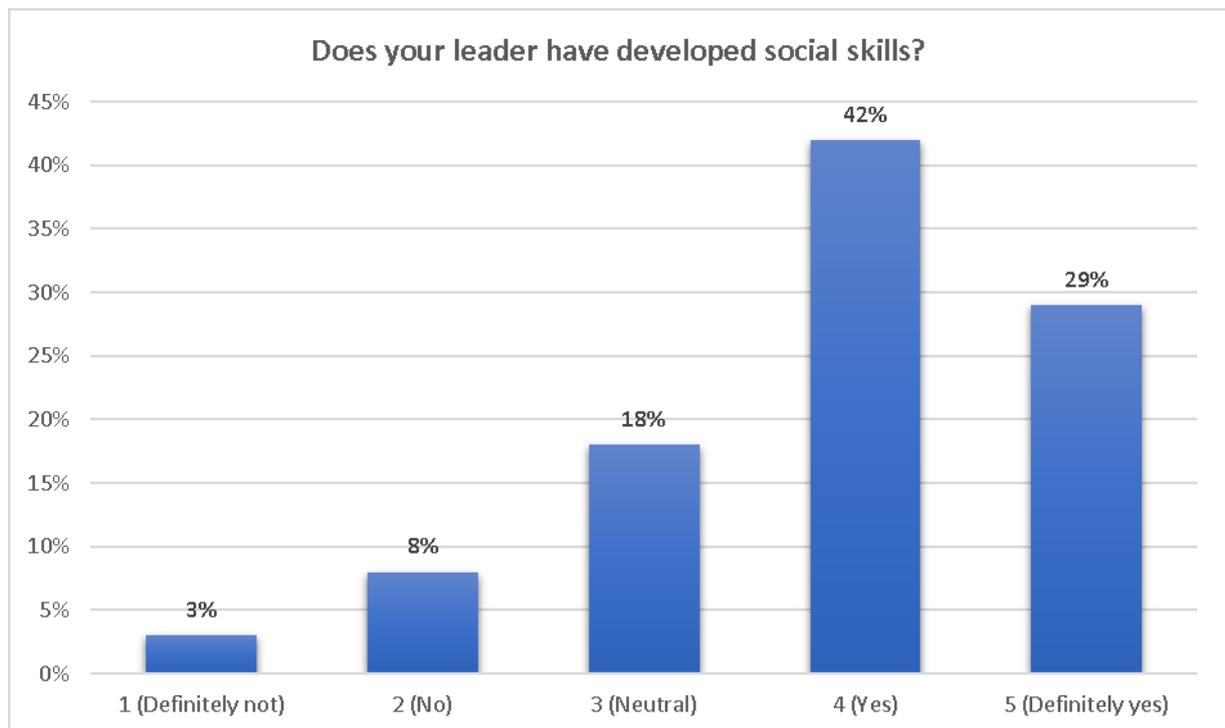

**Figure 14.** Answers from enterprise employees who have experienced a variety of leaders' management styles to question 14 of the research survey.

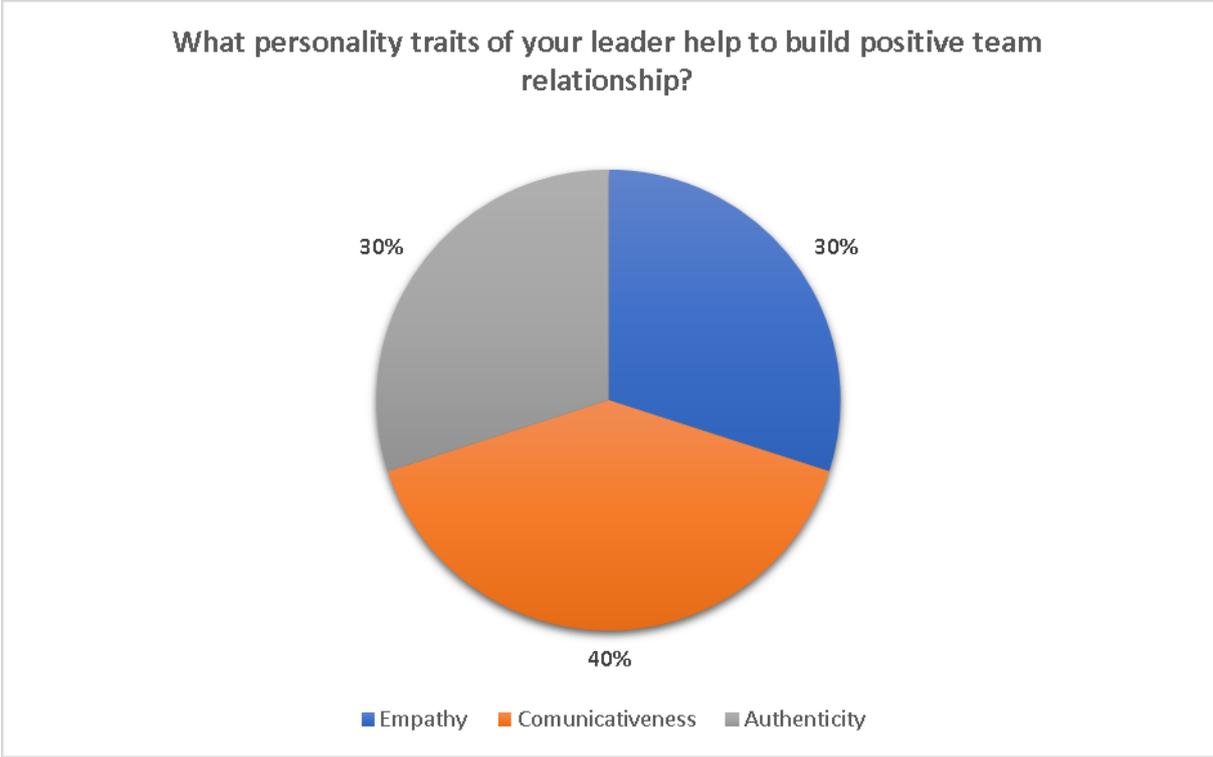

**Figure 15.** Answers from enterprise employees who have experienced a variety of leaders' management styles to question 15 of the research survey.

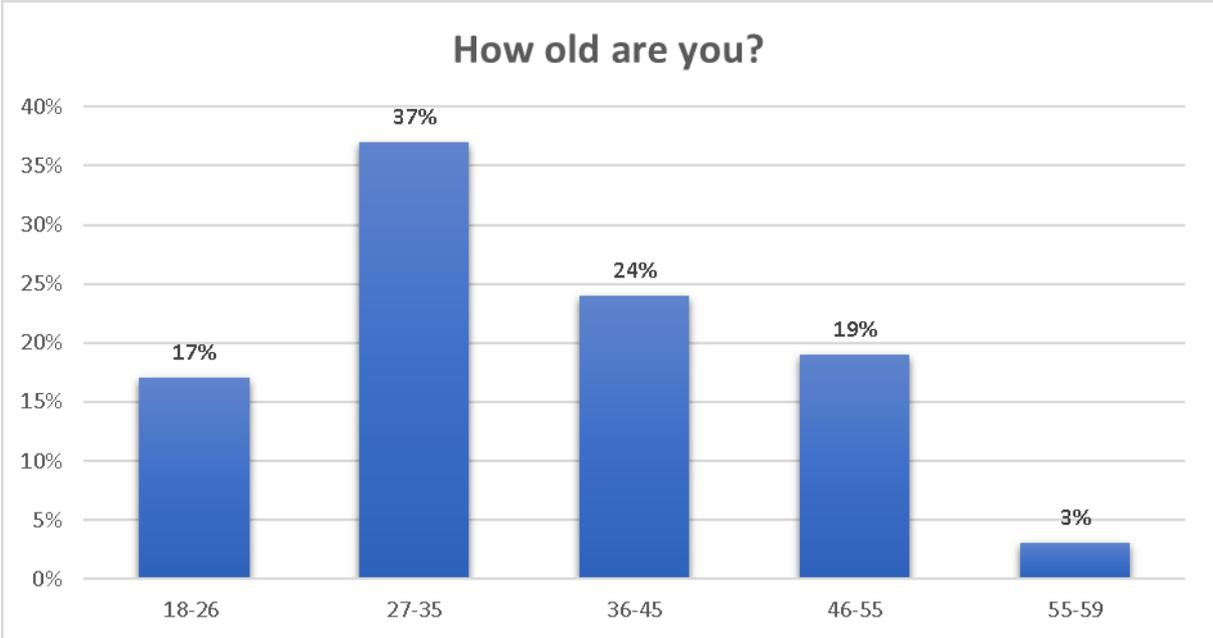

**Figure 16.** Answers from enterprise employees who have experienced a variety of leaders' management styles to question 16 of the research survey.

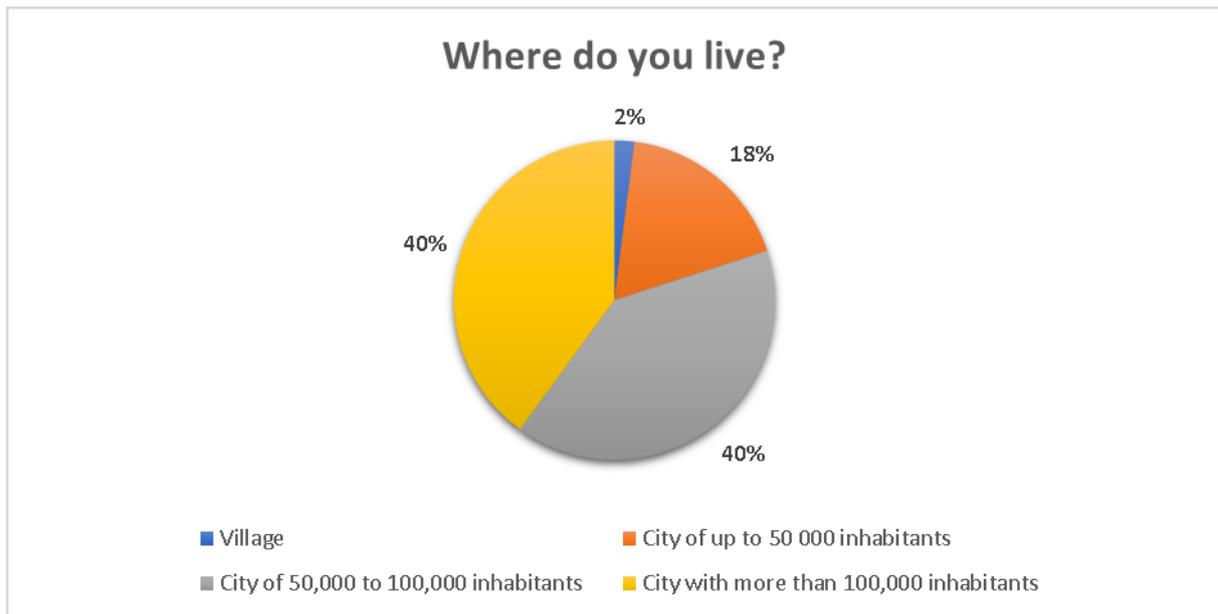

**Figure 17.** Answers from enterprise employees who have experienced a variety of leaders' management styles to question 17 of the research survey.

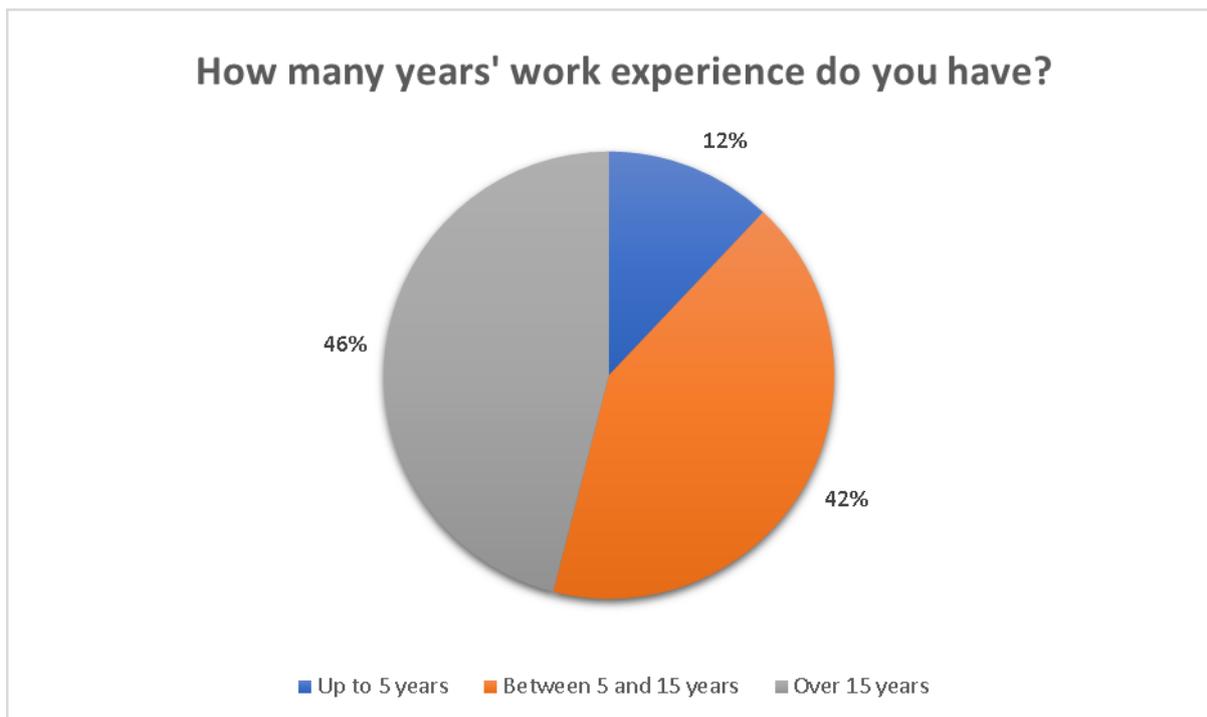

**Figure 18.** Answers from enterprise employees who have experienced a variety of leaders' management styles to question 18 of the research survey.

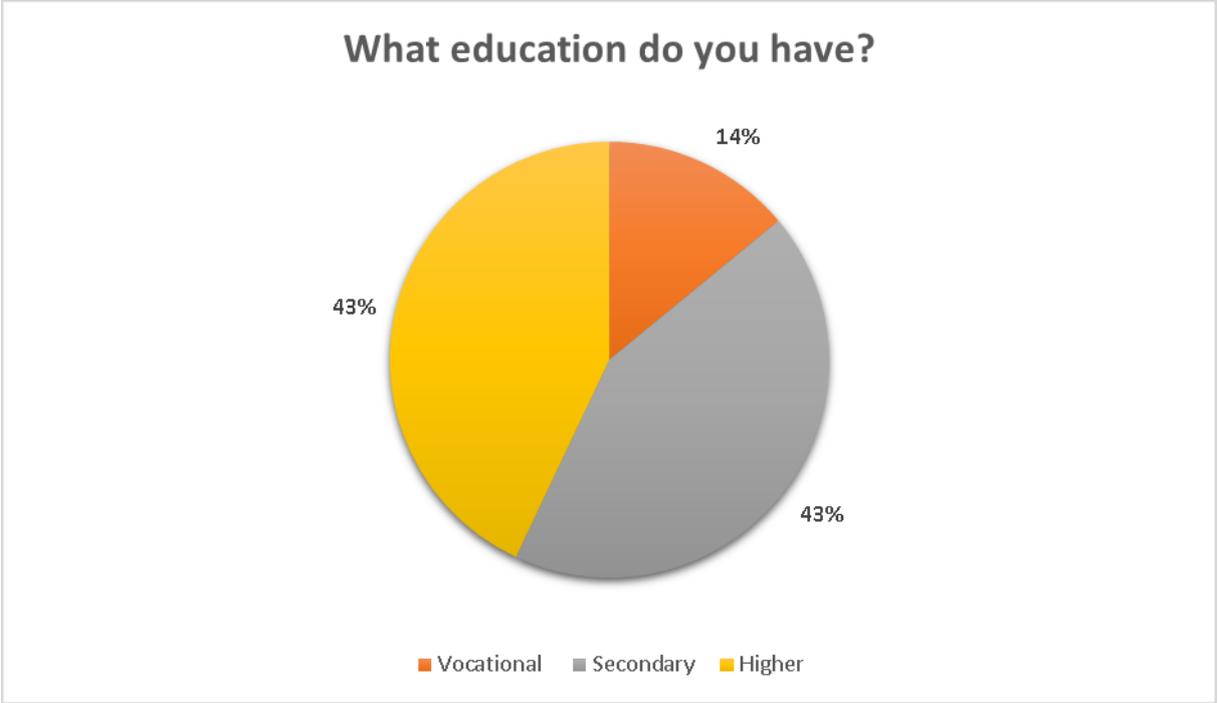

**Figure 19.** Answers from enterprise employees who have experienced a variety of leaders' management styles to question 19 of the research survey.

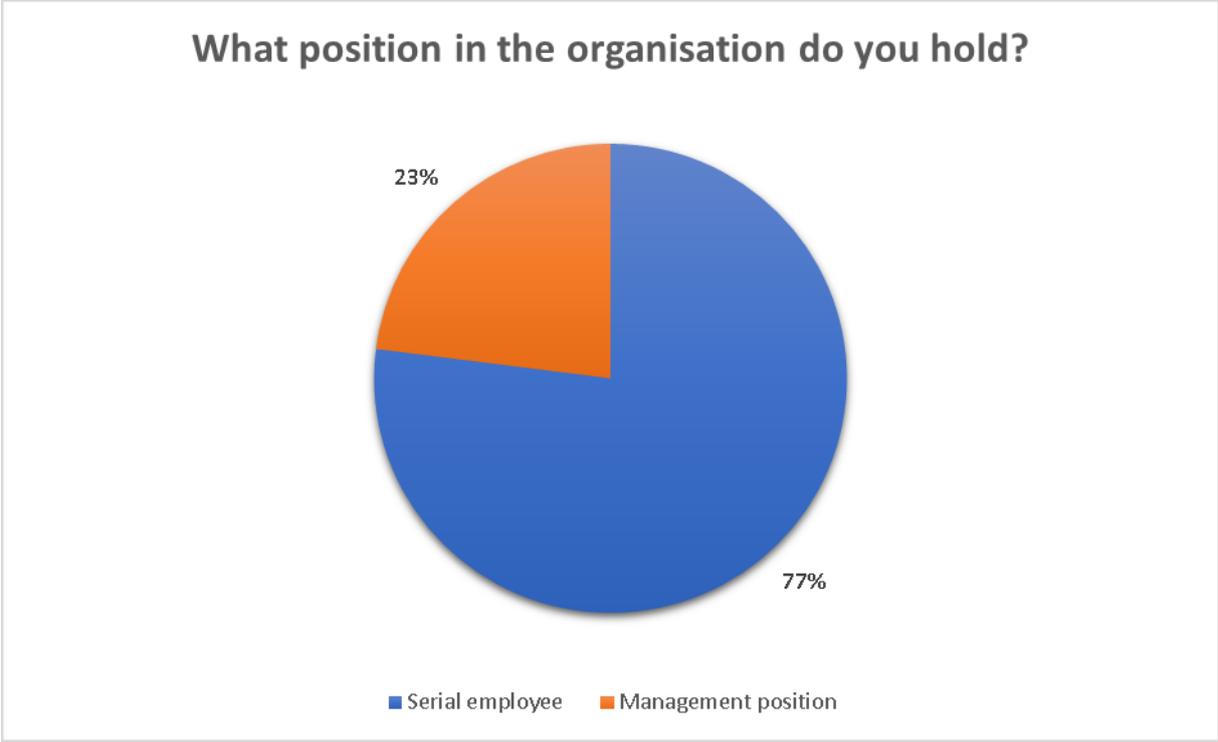

**Figure 20.** Answers from enterprise employees who have experienced a variety of leaders' management styles to question 20 of the research survey.

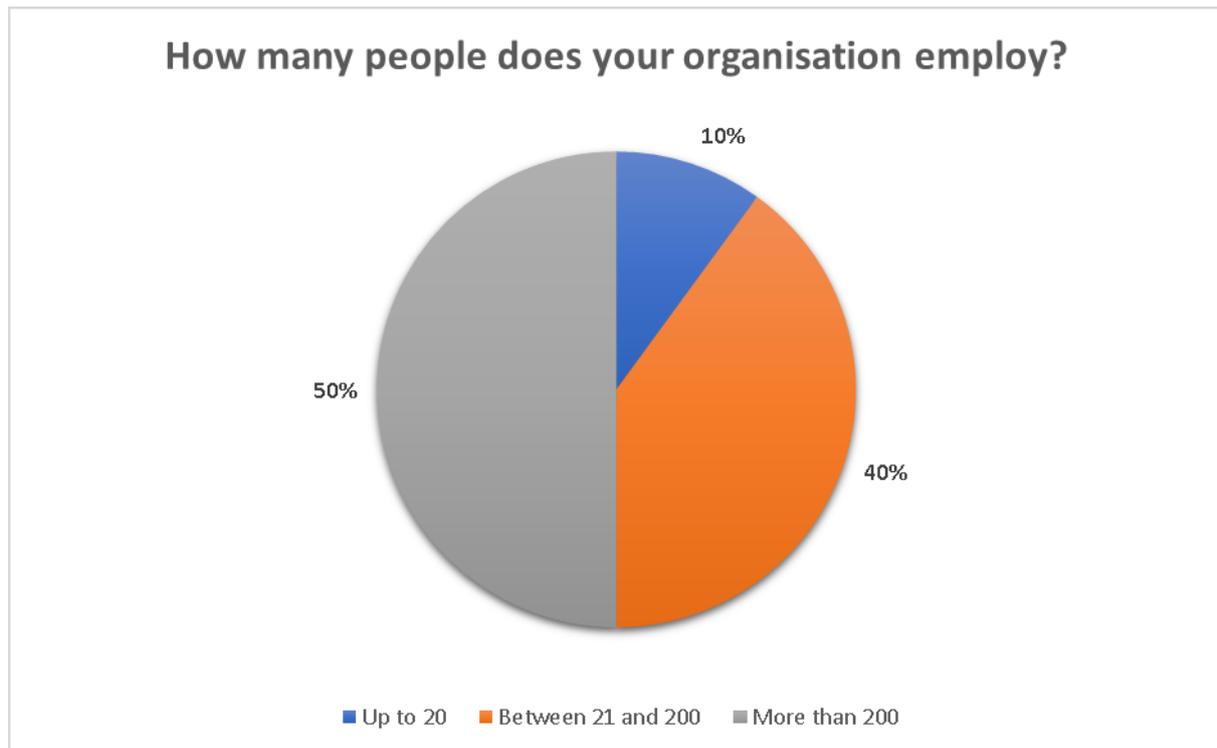

**Figure 21.** Answers from enterprise employees who have experienced a variety of leaders' management styles to question 21 of the research survey.

The analysis reveals that most respondents recognize emotional intelligence as a crucial factor in leadership effectiveness. The study employed elements from both Goleman's mixed model, which integrates personality traits, and Mayer and Salovey's ability model, focusing specifically on emotional competencies. Due to its pilot nature and limited sample size (n = 100), the results should be considered preliminary, indicating a significant role of emotional intelligence in team management but requiring further validation through larger and more diverse research samples.

## 4.2. Analysis of research results

The first question in the survey served as an introductory inquiry, assessing respondents' familiarity with the concept of emotional intelligence. The results indicate that the vast majority of participants (84%) confirmed their awareness of the term, whereas 16% admitted that they were unfamiliar with it.

The following question aimed to explore how respondents understood the concept of emotional intelligence. Participants were presented with three possible definitions and asked to choose one. The most frequently selected response (41%) defined emotional intelligence as the ability to control and manage one's emotions in stressful and conflict situations. A nearly equal proportion (40%) of respondents emphasized the ability to recognize and understand one's own emotions and those of others within a team. The least selected option (19%) linked emotional intelligence to decision-making based on data analysis and intuition.

The responses revealed varying assessments of emotional intelligence levels among leaders. A total of 1 respondent (1%) assigned a rating of 1, indicating a very low level of emotional intelligence. Similarly, 8 respondents (8%) rated their leader at level 2, suggesting a low level of emotional intelligence, albeit slightly higher than the previous category. A moderate level of emotional intelligence was reported by 21 respondents (21%), who selected a rating of 3. The largest proportion of participants, 45 respondents (45%), assigned a rating of 4, indicating a high level of emotional intelligence. Furthermore, 25 respondents (25%) explicitly stated that their leaders exhibited a high level of emotional intelligence. These findings suggest that a substantial majority of respondents perceive their leaders as possessing at least a moderate to high level of emotional intelligence, with a significant proportion assigning ratings at the upper end of the scale.

Respondents were asked about traits that confirm a leader's high emotional intelligence. Of the respondents, as many as 37% believe that an analytical mind is the most important indicator of high intelligence in a leader. 34% of respondents indicated that fairness best reflects a leader's high emotional intelligence. Empathy received the fewest responses (29%), suggesting that it is seen as a less important element of a leader's emotional intelligence.

An essential aspect of the study focused on assessing leaders' ability to manage their emotions during conflicts. The results indicate that over 80% of respondents perceive their leaders as capable of effectively regulating their emotions in such situations. Conversely, 19% of participants reported that they had not had the opportunity to observe these competencies in their leaders.

Additionally, the survey included a question evaluating team members' perceptions of their leaders' emotional expression. A majority of respondents (56%) indicated that their leader expresses emotions in a manner appropriate to the given situation. Meanwhile, 28% of participants assessed that their leader remains composed and does not outwardly display emotions. The smallest proportion of respondents (16%) perceived their leader as reacting nervously in conflict scenarios. These findings suggest that most leaders are viewed as demonstrating effective emotional management, with a prevailing tendency toward either appropriate emotional expression or emotional restraint.

Respondents were also asked about how their leaders resolve conflicts. The most frequently selected answer was that the leader conducts a dialogue with his subordinates (45%). A smaller number of respondents indicated that they had witnessed situations where their superiors waited for conflicts to resolve themselves (36%). The fewest respondents (19%) experienced stressful conversations between leaders and employees.

Respondents were asked to assess their leaders' empathy toward their subordinate team on a scale from 1 to 5, where 1 signifies "definitely not" and 5 denotes "definitely yes". The results indicate that 63% of respondents perceive their leaders as highly empathetic (rating 4 or 5), while 21% rate their leaders' empathy at an average level (3), and 16% believe their leaders lack empathy (rating 1 or 2). The statistical analysis further supports these findings, with a mean score of 3.78, reflecting a generally positive perception, and a median score of 4, indicating that at least half of the respondents rated their leaders' empathy at this level or higher. The standard deviation of 1.08 suggests moderate variability in responses, highlighting differing experiences among employees. These results confirm that a significant proportion of employees recognize empathy in their leaders.

As part of the survey, respondents were asked to evaluate the influence of leaders' empathy on team collaboration. The majority (62%) reported a positive impact, highlighting that empathetic leadership fosters team integration and enhances motivation. However, 19% of respondents noted a lack of team engagement, while an equal proportion (19%) indicated that insufficient empathy from leaders contributed to stress within the team. These findings suggest that while empathy is largely perceived as a crucial factor in strengthening teamwork, its absence may lead to adverse effects on team dynamics.

Another critical dimension of the study involved assessing the perception of ethical leadership. The data presented in the accompanying chart illustrate respondents' views on the extent to which their leaders adhere to ethical principles. A significant majority (74%) affirmed that their leaders demonstrate ethical behavior, reflecting a generally high assessment of their moral integrity and professional conduct. Conversely, 26% of respondents expressed concerns regarding the ethical practices of their leaders, suggesting the presence of ethical ambiguities or inconsistencies that warrant further investigation.

The next question was designed to detail the ethical actions of leaders. Respondents were asked to indicate what actions are taken by supervisors. 43% of respondents indicated that leaders' ethical actions are most evident in the area of treating co-workers. 30% of respondents believe that ethical actions are based on respecting co-workers and therefore: respecting the leader. The third group, which indicated the smallest number of responses, i.e. 27%, believes that ethical actions are evident in the way conflicts are resolved within the team.

Respondents were asked to rate the effectiveness of team motivation to achieve goals. The responses are in the form of a five-point scale, where 1 means that leaders are unable to motivate teams to any degree, while 5 means that leaders have the competence to motivate teams effectively. The vast majority of respondents, as many as 61%, believe that their leaders have the ability to effectively motivate a team to achieve goals. 28% of respondents rated their leaders as having moderate effectiveness, while 11% of responses indicate that leaders do not have any ability to motivate their subordinates to achieve their goals.

One of the key aspects examined in the survey was the identification of specific motivational strategies employed by leaders. The most frequently cited method, indicated by 45% of respondents, was the provision of professional development opportunities. Recognition and rewards from superiors were identified as a motivational factor by 33% of participants. The least frequently mentioned approach (22%) was the creation of a supportive and friendly work environment as a means of employee motivation. These findings suggest that while professional growth and recognition are primary drivers of motivation, workplace atmosphere is considered a less dominant factor.

Another critical dimension of the study involved assessing the level of social competence among leaders. Respondents evaluated their leaders' social skills on a five-point scale, where 1 represented a lack of developed social competencies and 5 denoted a high level of proficiency. The results indicate that over 70% of respondents perceive their leaders as possessing well-developed social competencies. Additionally, 18% of participants rated their leaders' social competence as moderate, while 11% expressed the view that their leaders exhibit a lack of social skills.

The 15 question was about the personality traits of leaders that have a positive impact in the process of building positive relationships in the team. Responses to this question are as follows: the largest group of respondents (40%) indicated that the leader's communicativeness has an impact on building positive relationships in the team. 30% felt that empathy was the most important trait. The same number of respondents (30%) believe that the leader's authenticity contributes to the creation of positive relationships in the team.

Further questions are metric questions. Participants in the survey are people of working age, which covers a wide age range. The youngest respondent participating in the survey is 19 years old, while the oldest is 59 years old. The largest number of respondents participating in the survey reside in cities while a minority (2%) live in villages. The largest group (43%) among those surveyed were respondents with higher education then those with secondary education. The vast majority (77%) of the respondents declared that they held positions as rank-and-file employees, while the rest held managerial positions. In terms of work experience, about half of the respondents declared that they have more than 15 years of education. Most people (50%) work in organisations that employ more than 200 people.

The correlation analysis examines, which is shown in figure 22 the relationship between key leadership characteristics, including emotional intelligence, empathy, motivation effectiveness, social competence, and ethical leadership. The correlation matrix provides insights into the strength and direction of these relationships. The heatmap visualizes these correlations, highlighting the strongest relationships in warm colors and weaker associations in cooler tones. These findings emphasize the interconnected nature of leadership traits and provide empirical support for leadership development strategies focusing on emotional intelligence, ethics, and interpersonal skills.

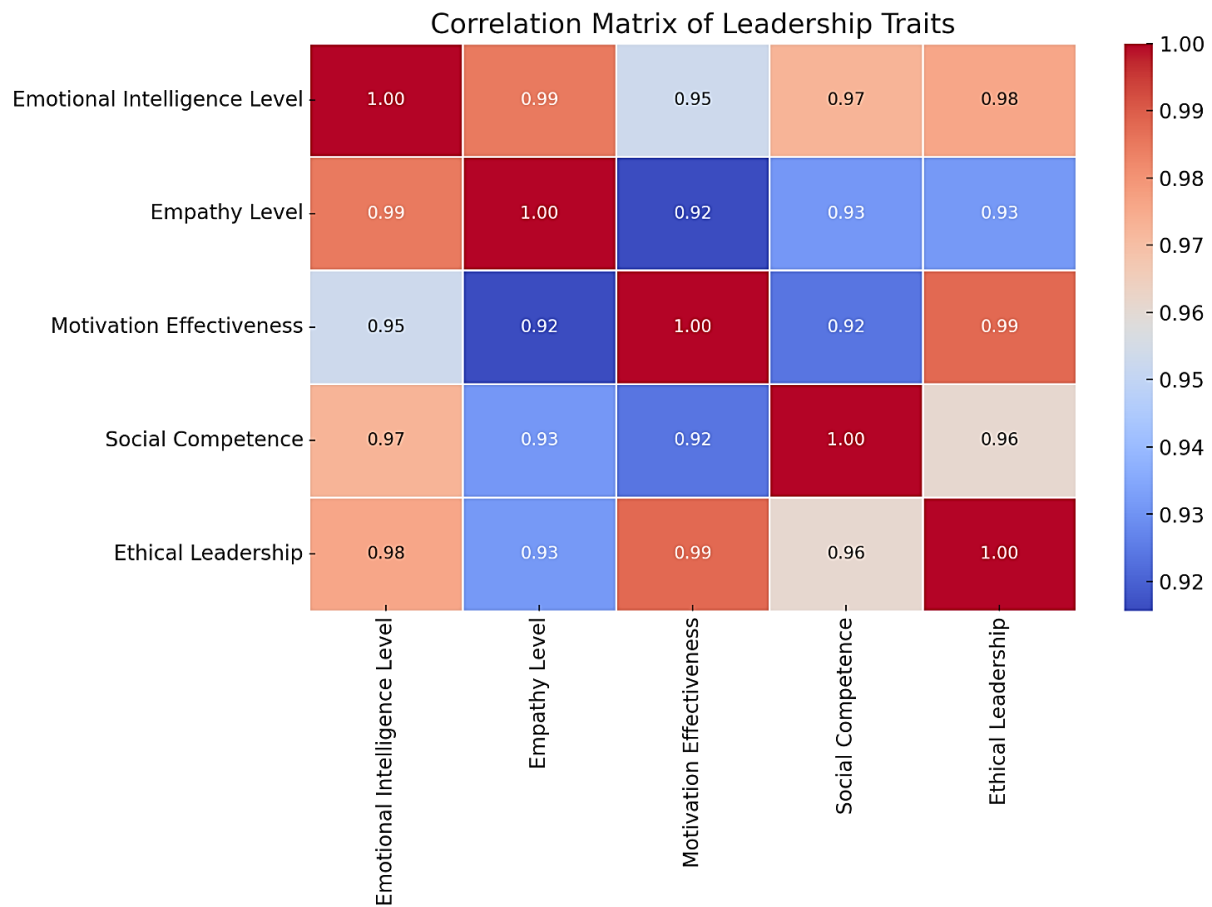

**Figure 22.** Correlation Matrix of Leadership Traits created from survey data.

The analysis revealed the following correlations:

Strong correlation between Emotional Intelligence and Empathy (0.98) - leaders with higher levels of emotional intelligence tend to be perceived as more empathetic. Emotional Intelligence and Social Competence correlation (0.97) - leaders with well-developed emotional intelligence are also rated highly in terms of social competence. This aligns with existing theories suggesting that emotional intelligence is a foundational component of effective interpersonal skills in leadership. Ethical Leadership and Motivation Effectiveness (0.99) - the strongest correlation in the dataset indicates that leaders perceived as ethical are also seen as highly effective in motivating their teams. Empathy and Ethical Leadership (0.93) - a high correlation suggests that leaders who are seen as empathetic are also perceived as ethical. Motivation Effectiveness and Social Competence (0.92) - social competence is crucial for motivating teams effectively. Leaders who can navigate social dynamics skillfully are better equipped to motivate their teams.

In summary, the correlation analysis reinforces the premise that key leadership traits do not operate in isolation but are closely interwoven, with emotional intelligence emerging as a central factor. The strongest associations, particularly between emotional intelligence, empathy, social competence, and ethical leadership, suggest that cultivating emotional awareness and interpersonal sensitivity significantly enhances a leader's effectiveness.

These findings provide practical guidance for leadership development programs, emphasizing the need to nurture emotional and ethical dimensions alongside technical competencies to foster holistic and impactful leadership.

## 5. Conclusions

The analysis of the collected empirical material provided important information on the emotional intelligence of leaders and its impact on team management. The research was conducted among 100 employees who had the opportunity to experience a variety of leaders' management styles at different stages of their careers. The collected data, including evaluations and opinions of the respondents, made it possible to draw conclusions.

The results of the study present that emotional intelligence is correlated with the positive perception of leaders by subordinates. The degree of development of emotional intelligence can be an evaluation criterion during recruitment processes and promotions.

The analysis confirms that emotional intelligence is a central factor in effective leadership, strongly correlating with empathy, social competence, and ethical leadership. Leaders who demonstrate emotional intelligence are also perceived as more ethical, empathetic, and socially competent, reinforcing the idea that emotional awareness enhances leadership effectiveness.

Additionally, a very strong correlation between ethical leadership and motivation effectiveness suggests that employees are more motivated when they perceive their leaders as ethical and fair. The findings highlight that social competence plays a key role in motivating teams, further emphasizing the importance of interpersonal skills in leadership.

Survey results show that leaders' ethical performance is highly valued by employees. As a result, companies can introduce a code of ethics for employees to promote leaders and their ethical actions and can implement business ethics training that can help leaders make decisions that are in line with the company's mission and values.

The results of our own research indicate the importance of emotional intelligence in managing subordinate personnel. Leaders who develop competence in this area contribute to the creation of more integrated, motivated and effective teams. In business practice, this means that organizations should invest in emotional intelligence training for their leaders. This is important to promote an organizational culture based on empathy, fairness and effective conflict resolution.

In order to strengthen the practical applicability of this study, it is recommended that organizations develop targeted emotional intelligence programs specifically for leaders and managers. Such programs may include workshops focusing on key interpersonal competencies such as active listening, empathy, emotional self-regulation, and stress management. Additionally, incorporating coaching methods tailored to enhancing emotional intelligence and integrating EI assessment tools, such as EI-focused 360-degree feedback instruments, into existing leadership evaluation systems would facilitate the practical implementation of emotional intelligence principles in managerial practice.

Considering the methodological limitations arising from the pilot nature of this study due to the limited number of respondents, it is recommended to conduct further research involving a larger and more diverse sample of specialists representing various sectors and organizational contexts. Such expanded research would enable broader validation of the obtained results, enhance their generalizability, and strengthen the practical implications of the study's conclusions.